\documentclass[switch]{article} 

\usepackage{preprint}

\usepackage{amsmath, amsthm, amssymb, amsfonts}

\usepackage[numbers,square]{natbib}
\usepackage[utf8]{inputenc}	
\usepackage[T1]{fontenc}	
\usepackage{xcolor}		
\usepackage[hidelinks]{hyperref}	
\usepackage{booktabs} 		
\usepackage{nicefrac}		
\usepackage{microtype}		
\usepackage{lineno}		
\usepackage{float}			
\usepackage{graphicx}
\usepackage{lipsum}		

\usepackage{newfloat}
\DeclareFloatingEnvironment[name={Supplementary Figure}]{suppfigure}
\usepackage{sidecap}
\sidecaptionvpos{figure}{c}

\usepackage{titlesec}
\titlespacing\section{0pt}{12pt plus 3pt minus 3pt}{1pt plus 1pt minus 1pt}
\titlespacing\subsection{0pt}{10pt plus 3pt minus 3pt}{1pt plus 1pt minus 1pt}
\titlespacing\subsubsection{0pt}{8pt plus 3pt minus 3pt}{1pt plus 1pt minus 1pt}

\title{Structure and Design of HoloGen}

\usepackage{xwatermark}
\newwatermark[allpages,color=gray!60,angle=90,scale=0.32, xpos=-4.05in,ypos=0]{\href{https://gitlab.com/CMMPEOpenAccess/HoloGen}{\color{gray}{HoloGen v2.2.1.17177}}}
\newwatermark[allpages,color=gray!60,angle=90,scale=0.32, xpos=3.9in,ypos=0]{\href{https://gitlab.com/CMMPEOpenAccess/HoloGen}{\color{gray}{\today}}}
\newwatermark[allpages,color=gray!90,angle=0,scale=0.28, xpos=0in,ypos=-5in]{pjc209@cam.ac.uk}

\usepackage{authblk}

\author[1]{Peter J. Christopher}
\author[1]{Timothy D. Wilkinson}
\affil[1]{Centre of Molecular Materials, Photonics and Electronics, University of Cambridge, UK}

\begin{document}

        \maketitle
        
        \begin{abstract}
            Increasing popularity of augmented and mixed reality systems has seen a similar increase of interest in 2D and 3D computer generated holography (CGH). Unlike stereoscopic approaches, CGH can fully represent a light field including depth of focus, accommodation and vergence. Along with existing telecommunications, imaging, projection, lithography, beam shaping and optical tweezing applications, CGH is an exciting technique applicable to a wide array of photonic problems including full 3D representation. 
            
            Traditionally, the primary roadblock to acceptance has been the significant numerical processing required to generate holograms requiring both significant expertise and significant computational power. 
            
            This article discusses the structure and design of HoloGen. HoloGen is an MIT licensed application that may be used to generate holograms using a wide array of algorithms without expert guidance. HoloGen uses a Cuda C and C++ backend with a C$\#$ and Windows
            Presentation Framework graphical user interface. The article begins by introducing HoloGen before providing an in-depth discussion of its design and structure. Particular focus is given to the communication, data transfer and algorithmic aspects.
        \end{abstract}
        \vspace{0.35cm}

    \normalsize
    
    \section{Introduction}
    
    This article presents an overview of the structure of the HoloGen application as well as some of the design decisions made during development. While the majority of HoloGen is simple contextually and requires little discussion, a few areas stand out as meriting explanation for future developers. Key features for further discussion include: The parameter and command tree in Section~\ref{codediscuss1}; The native algorithm interface in Section~\ref{codediscuss2}; The serialisation architecture in Section~\ref{codediscuss3} and the user interface construction in Section~\ref{codediscuss4}. Section~\ref{codediscuss5} discusses the role of all the application libraries and their connections as well as any key classes not covered elsewhere. Finally Section~\ref{codediscuss6} provides some areas of future expansion and the infrastructure put in place to handle them.
    
    \section{HoloGen} \label{hololicenses} \label{HoloGen}
    
    The HoloGen Suite is an application suite built on top of a custom parameter framework in C$\#$ and WPF. The package is open-source under the MIT license with sub libraries each having their own license. The source is available online.\footnote{\url{https://gitlab.com/CMMPEOpenAccess/HoloGen}} Version 2.2.1.17177 of HoloGen runs to $~76,000$ lines of code with dependencies on 25 external libraries.
    
    \begin{figure}[tbhp]
        \centering
        
        {\includegraphics[width=0.6\linewidth,page=4]{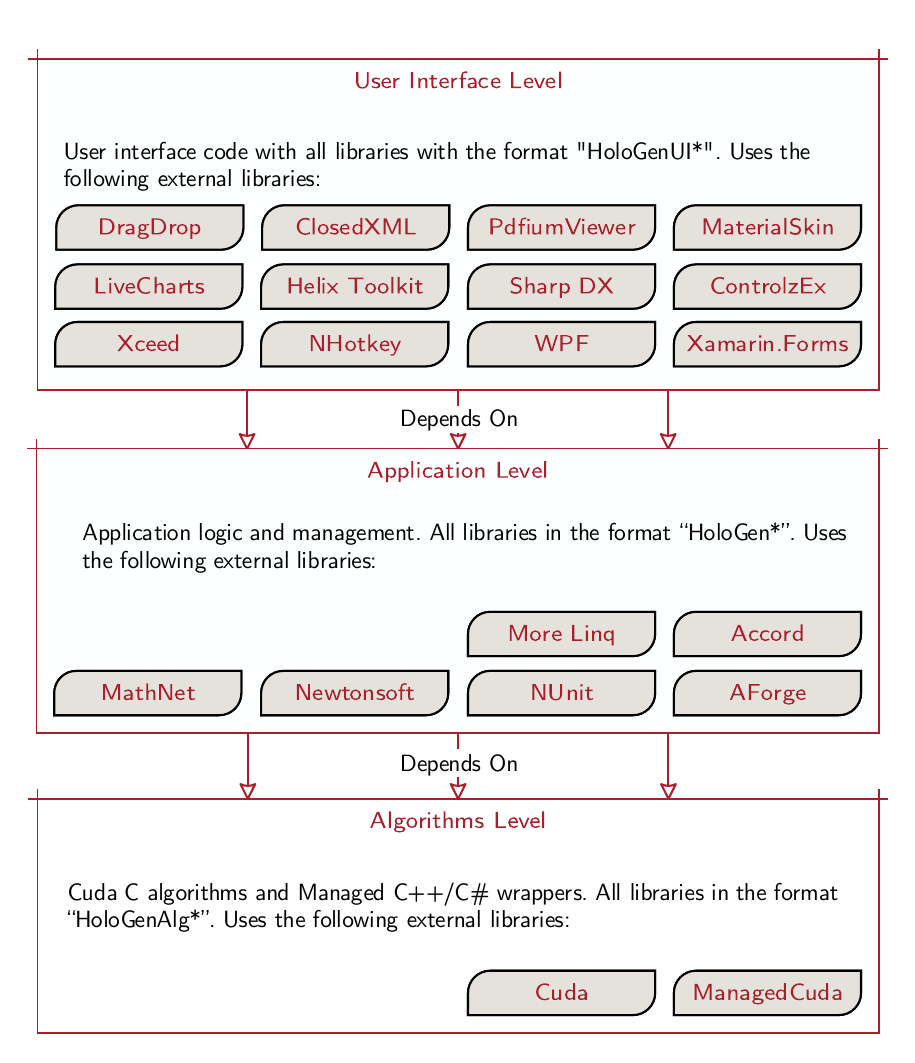}}
        
        \caption[HoloGen Application Levels]{HoloGen application levels. c.f. Figures~\ref{fig:codelayout2a}, \ref{fig:codelayout2b} and \ref{fig:codelayout2c}. }
        \label{fig:codelayout1}
    \end{figure}
    
\begin{figure}[tbhp]
           \centering
           {\includegraphics[width=1.0\linewidth,page=1]{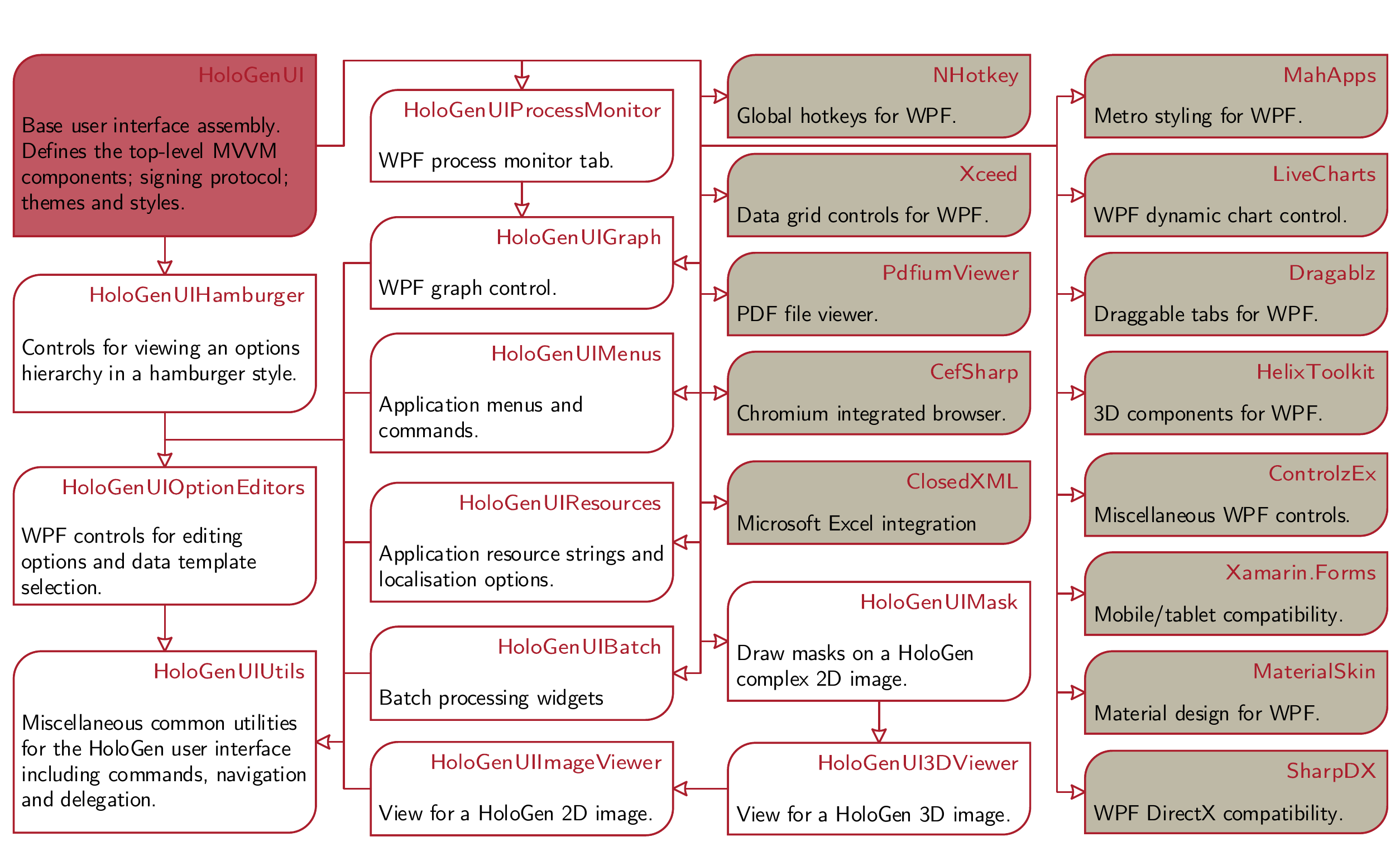}}
           \caption[HoloGen User Interface Level]{HoloGen user interface level. c.f. Figure~\ref{fig:codelayout1}. }
           \label{fig:codelayout2a}
       \end{figure}

    The HoloGen application is built on a MVVMA architecture. This is a standard Model-View-ViewModel~(MVVM) framework common in C$\#$ Windows Presentation Framework~(WPF) applications with an additional algorithms level written in a more traditional procedural/functional style on top of an Nvidia Cuda architecture interfaced in C++~\cite{model10}. This translates into a layered application structure as shown in Figure~\ref{fig:codelayout1}. This shows the three application levels: the user interface level; application level and algorithm level. These all depend only on levels beneath them and have their own independent imported libraries. A more detailed breakdown is shown in Figures~\ref{fig:codelayout2a}, \ref{fig:codelayout2b} and \ref{fig:codelayout2c}.  
    
    The cuFFT library from NVidia is used to perform the FFT element of the algorithms due to its high performance~\cite{steinbach2017gearshifft}. This is built on top of a Cuda framework with Thrust wrapper. Previous researchers have also used the FFTW library~\cite{Shimobaba2009,Shimobaba2012}, OpenCV~\cite{Wang2018}, OpenMP~\cite{Takada2012}, the Computational Wave Optics (CWO)~\cite{Shimobaba2017,Shimobaba2018a,Shimobaba2018} and Intel Math Kernel (MKL)~\cite{Matsushima2009} libraries as well as custom implementations in Matlab~\cite{Kim2008,Kim2009,Jia2013,Makey2013,Memmolo2014,Wang2019}, C/C++~\cite{Carpenter2010,Takada2012,Matsushima2009,Shimobaba2009,Shimobaba2012, Shimobaba2017,Shimobaba2018a,Shimobaba2018} and Python.
    
    HoloGen depends on a number of third-party libraries~\cite{model100, model101, model102, model103, model104, model105, model106, model107, model108, model109, model110,model130,model131, model132, model133, model400, cheese01, cheese03}. Most are licensed under the MIT license~\cite{model200, model201, model202, model203, model204, model206, model207, model208, model12s2, model131,model401} with several licensed under the Lesser General Public License~\cite{model209, model210,model230, model13}, several under the Apache License 2.0~\cite{model205, model232, model11, cheese01} and two under the three-clause BSD license~\cite{model233, model233a, cheese04}. These are all highly permissive licenses. Xamarin.Forms, C$\#$ and WPF come with appropriate usage licenses as part of Cambridge University Visual Studio package while Cuda comes with an appropriate license and EULA~\cite{model14, model15}.
    
    HoloGen includes a number of novel elements. A reflection based parameter framework allows for persistent parameter models with limited code reuse. A dynamic module system interfaces with Cuda implemented algorithms allowing for real-time Cuda compilation on host machines improving performance and utilising more advanced features of  newer graphics cards. three dimensional visualisation techniques can be used while viewing the generated hologram statistics. Fourier transform functionality is incorporated directly into the viewer. A tabular batch processing framework allows for multiple operations to be scheduled for background operation. Advanced tabulation allows for comparison of different holograms. New image and file types are introduced to handle the additional information available and all results are tagged with parameter metadata used to ensure traceability~\cite{blinder2019signal}. Provisional translations into French, Spanish, German and Chinese are available as well as complete help documentation. An included Chromium browser allows for integrated reporting features.       
       
    \section{Parameter and Command Hierarchy}\label{codediscuss1}
    
    HoloGen uses a custom reflection based parameter and command system. This is in contrast to the XML parameter sheet systems widely in use. 
    Instead of the parameter types and interactions being defined in parameter sheets which are parsed at runtime, the parameter system
    is coded into the C$\#$ directly. This significantly reduces the runtime overhead as well as improves the error checking available 
    at compile time. The downside is an increased architecture exposure of the parameter hierarchy. This structure is contained in the \textit{HoloGenHierarchy} library. The decision to use the reflection based system was made to enable a side project of the author.
    
    \subsection{Class Inheritance Hierarchy}
    
    Two distinct areas of the \textit{HoloGenHierarchy} library code stand out: the parameter/command types (e.g. numerical options, menu commands, etc) and the tree of elements containing them (e.g. pages, menus, etc).
    
    The parameter types are defined using the following key classes shown in Figure~\ref{fig:hierarch1}: 
    
    \begin{itemize}
        \item \hypertarget{INode}{\texttt{INode}} 
        - Base interface for all \textit{leaf} nodes in the parameter hierarchy. 
        
        \item \hypertarget{ICommand}{\texttt{ICommand}} 
        - Base command interface extended by all parameter types. Extends \hyperlink{INode}{\texttt{INode}}.
        
        \item \hypertarget{IOption}{\texttt{IOption}} 
        - Base parameter interface extended by all parameter types. Extends \hyperlink{INode}{\texttt{INode}}.
        
        \item \hypertarget{Command}{\texttt{Command}} 
        - Abstract base command class extended by all command types. Distinct from \hyperlink{ICommand}{\texttt{ICommand}} which it implements due to C$\#$ not handling generic template references.
        
        \item \hypertarget{Option}{\texttt{Option}} 
        - Abstract base parameter class extended by all parameter types. Distinct from \hyperlink{IOption}{\texttt{IOption}} which it implements due to C$\#$ not handling generic template references. A template system allows for generic manipulation of wrapped values without exposing the internals of the class to extending objects.
        
        \item \hypertarget{NumericOption}{\texttt{NumericOption}} 
        - Abstract base parameter class extended by all numeric parameter types. Extends \hyperlink{Option}{\texttt{Option}}.
        \item \hypertarget{IntegerOption}{\texttt{IntegerOption}} 
        - Integral numeric parameter type. Extends \hyperlink{NumericOption}{\texttt{NumericOption}}.
        
        \item \hypertarget{DoubleOption}{\texttt{DoubleOption}} 
        - Floating point numeric parameter type. Extends \hyperlink{NumericOption}{\texttt{NumericOption}}.
        
        \item \hypertarget{ILargeOption}{\texttt{ILargeOption}} 
        - Interface that flags to the display that implementing options require extra space on the UI. Extends \hyperlink{IOption}{\texttt{IOption}}.
        
        \item \hypertarget{TextOption}{\texttt{TextOption}} 
        - Text based parameter type. Extends \hyperlink{Option}{\texttt{Option}}.
        
        \item \hypertarget{LargeTextOption}{\texttt{LargeTextOption}} 
        - Larger version of \hyperlink{TextOption}{\texttt{TextOption}}. Implements \hyperlink{ILargeOption}{\texttt{ILargeOption}}.
        
        \item \hypertarget{PathOption}{\texttt{PathOption}} 
        - Alternative to \hyperlink{TextOption}{\texttt{TextOption}} that handles file paths. Implements \hyperlink{ILargeOption}{\texttt{ILargeOption}}.
        
        \item \hypertarget{ISelectOption}{\texttt{ISelectOption}} 
        - Base parameter interface extended by all selection based parameter types. Extends \hyperlink{IOption}{\texttt{IOption}}.

        \begin{figure}[tbhp]
            \centering
            {\includegraphics[width=\linewidth,page=2]{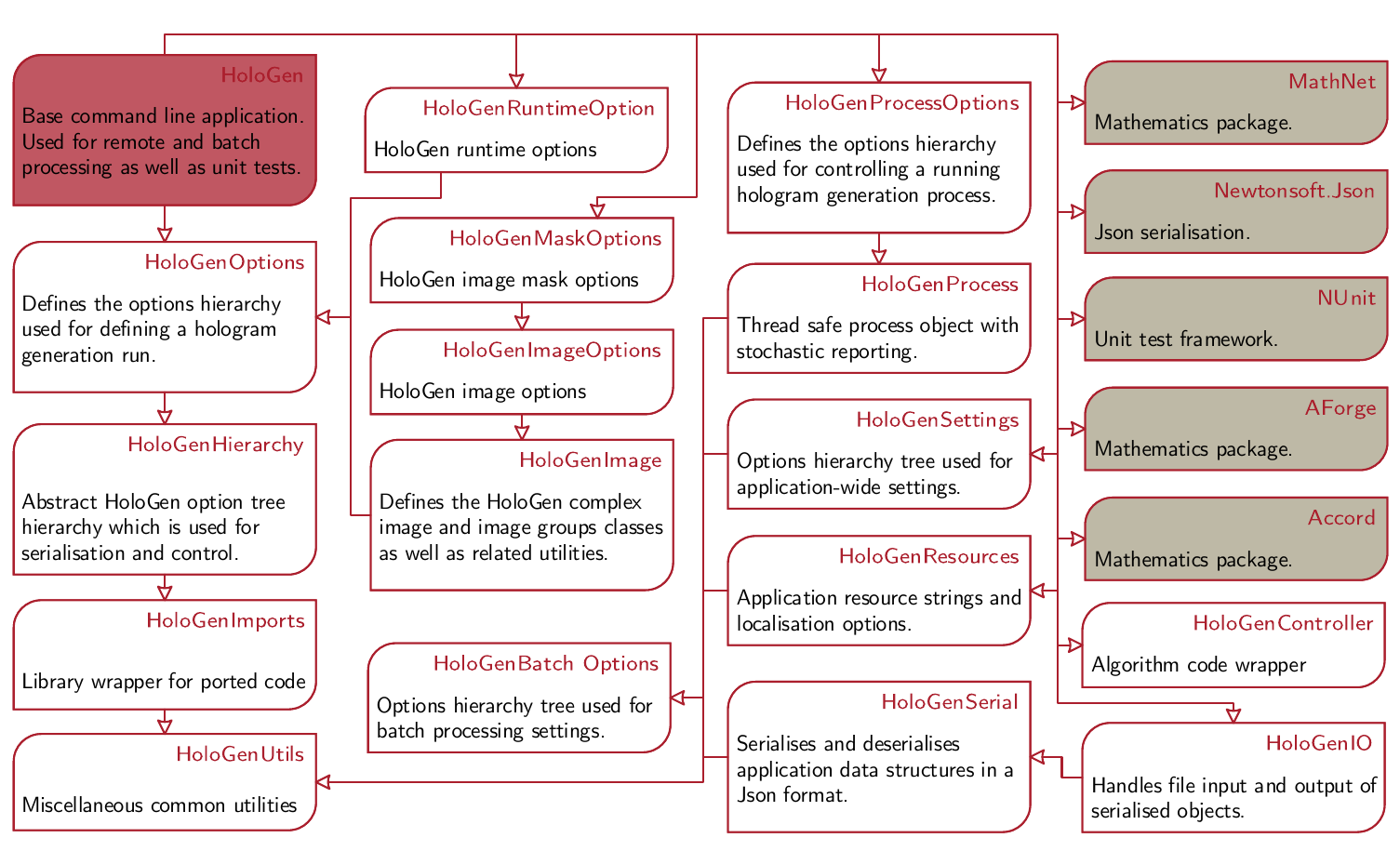}}
            \caption[HoloGen Application Level]{HoloGen application level. c.f. Figure~\ref{fig:codelayout1}. }
            \label{fig:codelayout2b}
        \end{figure}
        
        \item \hypertarget{SelectOption}{\texttt{SelectOption}} 
        - Abstract base parameter class extended by all selection based parameter types. Distinct from \hyperlink{ISelectOption}{\texttt{ISelectOption}} which it implements due to C$\#$ not handling generic template references. A template system allows for generic manipulation of wrapped values without exposing the internals of the class to extending objects. Contains a \hyperlink{PossibilityCollection}{\texttt{PossibilityCollection}} of \hyperlink{Possibility}{\texttt{Possibilities}} that can be selected as the option value. Any options owned by the selected \hyperlink{Possibility}{\texttt{Possibility}} are injected into the owning \hyperlink{HierarchyFolder}{\texttt{HierarchyFolder}}.
        
        \item \hypertarget{ListOption}{\texttt{ListOption}} 
        - Base class for parameters representing lists of values. Extends \hyperlink{Option}{\texttt{Option}}. Implements \hyperlink{ILargeOption}{\texttt{ILargeOption}}.
        
        \item \hypertarget{PathListOption}{\texttt{PathListOption}} 
        - Base class for parameters representing lists of files. Extends \hyperlink{Option}{\texttt{ListOption}}. Implements \hyperlink{ILargeOption}{\texttt{ILargeOption}}.
        
        \item \hypertarget{BooleanOption}{\texttt{BooleanOption}} 
        - Boolean value based parameter type. Extends \hyperlink{Option}{\texttt{Option}}.
        
        \item \hypertarget{BooleanOptionWithChildren}{\texttt{BooleanOptionWithChildren}} 
        - Boolean value based parameter type. Extends \hyperlink{BooleanOption}{\texttt{BooleanOption}}. When set to \textit{true}, any child options are injected into the owning \hyperlink{HierarchyFolder}{\texttt{HierarchyFolder}} in a manner similar to \hyperlink{SelectOption}{\texttt{SelectOption}}.
    \end{itemize}
    
\begin{figure}[tb]
        \centering
            {\includegraphics[width=0.7\linewidth,page=7]{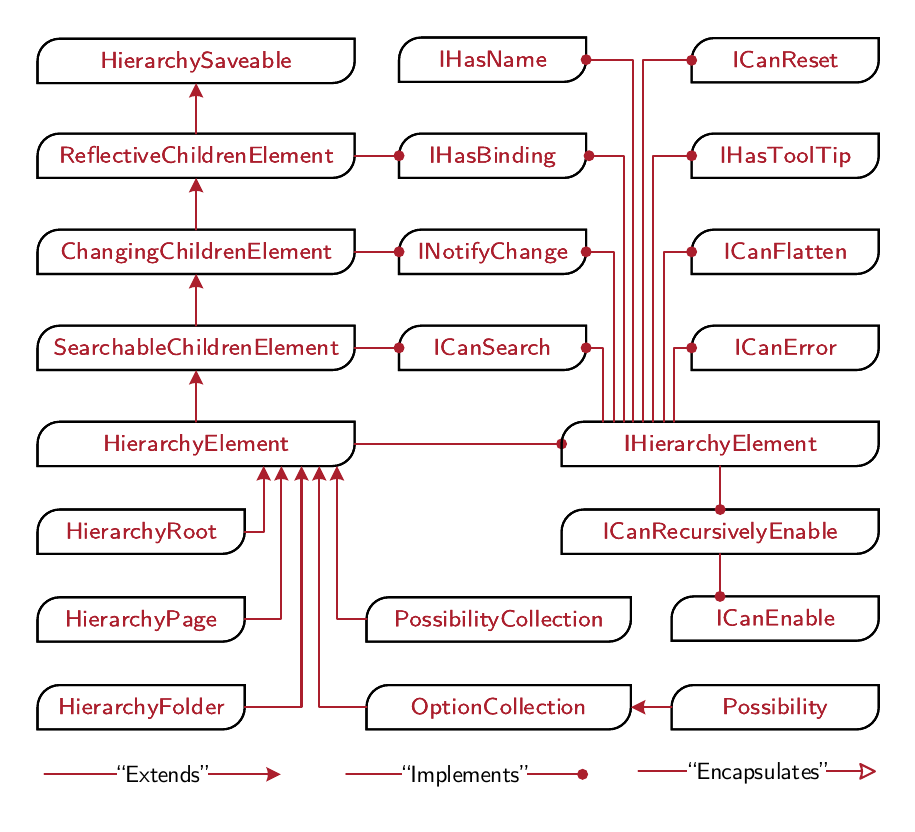}}
        \caption{HoloGen paramater tree inheritance hierarchy}
        \label{fig:hierarch2}
    \end{figure}

    The tree of elements containing the parameters or commands are shown in Figure~\ref{fig:hierarch2}:
    
    \begin{itemize}
        \item \hypertarget{ReflectiveChildrenElement}{\texttt{ReflectiveChildrenElement}} 
        - Key class representing any \textit{non-leaf} node in the parameter/command tree. When extended, this class uses the C$\#$ reflection system to find all public parameters of the same type as the specified template type and uses them to populate a list of children of that type. This allows extending classes to declare member elements without having to handle their manipulation or access.
        
        \item \hypertarget{ChangingChildrenElement}{\texttt{ChangingChildrenElement}} 
        - Extends \hyperlink{ReflectiveChildrenElement}{\texttt{ReflectiveChildrenElement}} to handle a \textit{non-leaf} node that has a changing set of children with appropriate notifications.
        
        \item \hypertarget{ChangingChildrenElement}{\texttt{ChangingChildrenElement}} 
        - Extends \hyperlink{ChangingChildrenElement}{\texttt{ChangingChildrenElement}} to handle a \textit{non-leaf} node that has a set of children which can be searched.
        
        \item \hypertarget{IHierarchyElement}{\texttt{IHierarchyElement}} 
        - Abstract interface for all \textit{non-leaf} nodes in the parameter tree.
        
        \item \hypertarget{HierarchyElement}{\texttt{HierarchyElement}} 
        - Abstract base class for all \textit{non-leaf} nodes in the parameter/command tree. Distinct from \hyperlink{IHierarchyElement}{\texttt{IHierarchyElement}} which it implements due to C$\#$ not handling generic template references.
        
        \begin{figure}[tbhp]
            \centering
            {\includegraphics[width=1.0\linewidth,page=3]{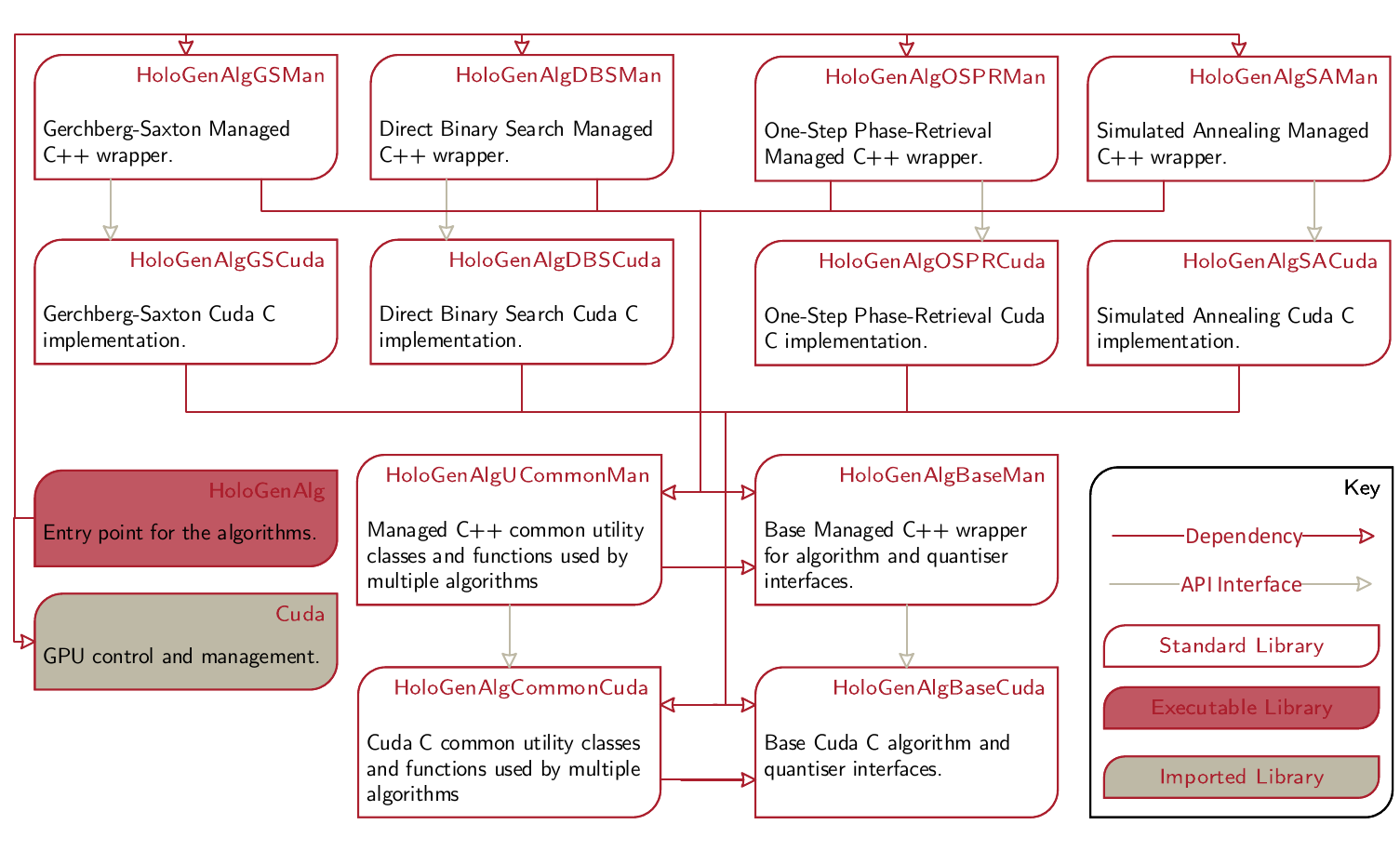}}
            \caption[HoloGen Algorithms Level]{HoloGen algorithms level. c.f. Figure~\ref{fig:codelayout1}. }
            \label{fig:codelayout2c}
        \end{figure}
        
        \item \hypertarget{HierarchyRoot}{\texttt{HierarchyRoot}} 
        - Implementation of \hyperlink{HierarchyElement}{\texttt{HierarchyElement}} that represents the root node of a parameter/command tree. Equivalent to the \textit{tab} or \textit{menu pop-out} level within HoloGen.
        
        \item \hypertarget{HierarchyPage}{\texttt{HierarchyPage}} 
        - Implementation of \hyperlink{HierarchyElement}{\texttt{HierarchyElement}} that represents a node of a parameter/command tree. Equivalent to the \textit{page} or \textit{menu} level within HoloGen.
        
        \item \hypertarget{HierarchyFolder}{\texttt{HierarchyFolder}} 
        - Implementation of \hyperlink{HierarchyElement}{\texttt{HierarchyElement}} that represents a node of a parameter/command tree. Equivalent to the \textit{folder} or \textit{sub-menu} level within HoloGen.
        
        \item \hypertarget{OptionCollection}{\texttt{OptionCollection}} 
        - Implementation of \hyperlink{HierarchyElement}{\texttt{HierarchyElement}} that represents a set of \hyperlink{Option}{\texttt{Options}} within a specialisation \hyperlink{Option}{\texttt{Option}} such as  \hyperlink{BooleanOptionWithChildren}{\texttt{BooleanOptionWithChildren}} or \hyperlink{SelectOption}{\texttt{SelectOption}}.
        
        \item \hypertarget{PossibilityCollection}{\texttt{PossibilityCollection}} 
        - Implementation of \hyperlink{HierarchyElement}{\texttt{HierarchyElement}} that represents a set of \hyperlink{Possibility}{\texttt{Possibilities}} within a \hyperlink{SelectOption}{\texttt{SelectOption}}.
        
        \item \hypertarget{Possibility}{\texttt{Possibility}} 
        - Represents a possibility state for a \hyperlink{SelectOption}{\texttt{SelectOption}}.
        
        \item \hypertarget{HierarchyVersion}{\texttt{HierarchyVersion}} 
        - Represents a version number for a parameter/command hierarchy.
        
        \item \hypertarget{HierarchySaveable}{\texttt{HierarchySaveable}} 
        - Represents any class that can be saved using the JSON serialisation.
    \end{itemize}
    
    In addition a number of function interfaces are used to mark exhibited behaviours at different levels of the hierarchy.
    
    \begin{itemize}
        \item \hypertarget{ICanEnable}{\texttt{ICanEnable}} 
        - Implemented by any class/interface that exhibits enable/disable behaviours.
        
        \item \hypertarget{ICanError}{\texttt{ICanError}} 
        - Implemented by any class/interface that can be in an error state.
        
        \item \hypertarget{ICanImportFromString}{\texttt{ICanImportFromString}} 
        - Implemented by any class/interface that allows for deserialisation from a string object.
        
        \item \hypertarget{ICanExportToString}{\texttt{ICanExportToString}} 
        - Implemented by any class/interface that allows for serialisation to a string object.
        
        \item \hypertarget{ICanFlatten}{\texttt{ICanFlatten}} 
        - Implemented by any \hyperlink{HierarchyElement}{\texttt{HierarchyElement}} that can flatten its internal tree structure.
        
        \item \hypertarget{IHasName}{\texttt{IHasName}} 
        - Implemented by any class/interface that has a name property.
        
        \item \hypertarget{IHasToolTip}{\texttt{IHasToolTip}} 
        - Implemented by any class/interface that has a tool tip property.
        
        \begin{figure}[tbhp]
            \centering
            
            {\includegraphics[width=0.7\linewidth,page=6]{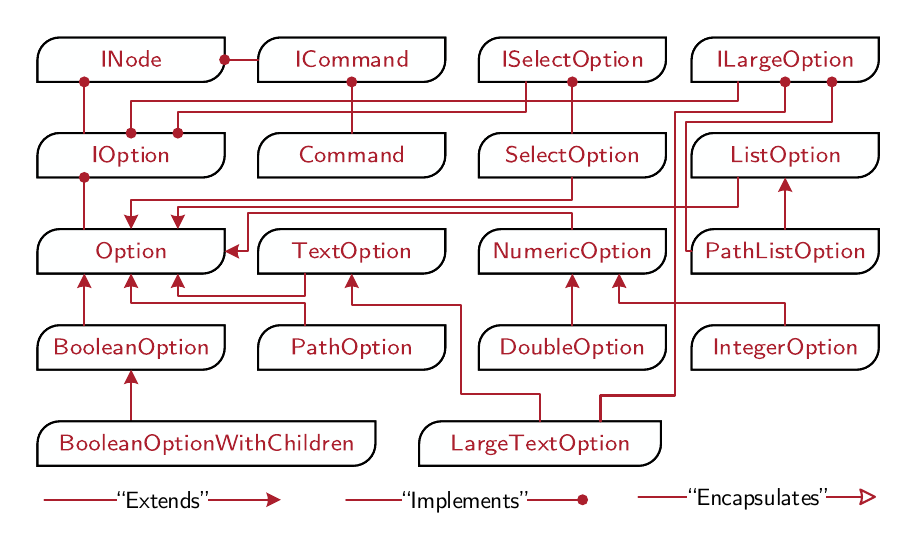}}
            
            \caption[HoloGen Parameter Types Inheritance Hierarchy]{HoloGen parameter types inheritance hierarchy }
            \label{fig:hierarch1}
        \end{figure}
        
        \item \hypertarget{IHasWatermark}{\texttt{IHasWatermark}} 
        - Implemented by any class/interface that has a watermark property.
        
        \item \hypertarget{IHasBindingPath}{\texttt{IHasBindingPath}} 
        - Implemented by any \hyperlink{HierarchyElement}{\texttt{HierarchyElement}}, \hyperlink{Command}{\texttt{Command}} or \hyperlink{Option}{\texttt{Option}} that the interface can be bound to given a link to the base of the parameter or command tree.
        
        \item \hypertarget{ICanSearch}{\texttt{ICanSearch}} 
        - Implemented by any class/interface that can be searched.
        
        \item \hypertarget{ICanReset}{\texttt{ICanReset}} 
        - Implemented by any class/interface that can be reset.
        
        \item \hypertarget{ICanRecursivelyEnable}{\texttt{ICanRecursivelyEnable}} 
        - Implemented by any class/interface that can set its own enabled state and that of its children. Extends \hypertarget{ICanEnable}{\texttt{ICanEnable}} .
        
        \item \hypertarget{INotifyChanged}{\texttt{INotifyChanged}} 
        - Implemented by any \hyperlink{Option}{\texttt{Option}} that notifies when the contained value changes.
    \end{itemize}
    
    \begin{figure}[tbhp]
        \centering
        
            {\includegraphics[width=0.7\linewidth,page=8]{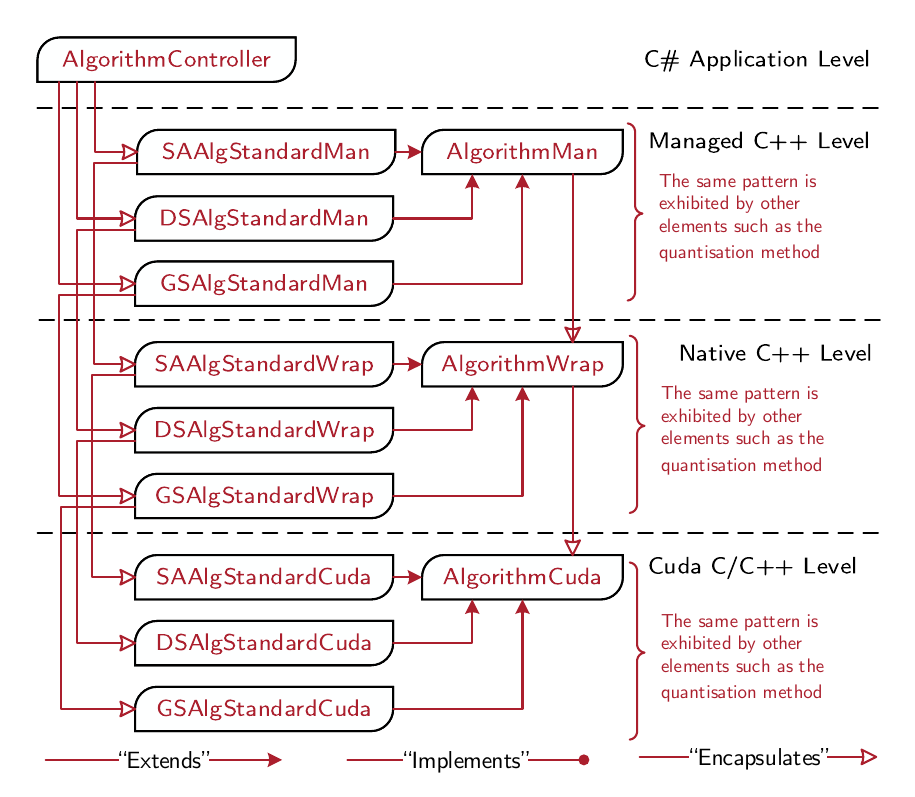}}
        
        \caption[HoloGen Algorithm Interface]{HoloGen algorithm interface }
        \label{fig:hierarch3}
    \end{figure}

    \subsection{Application}
    
    HoloGen defines a number of parameter and command hierarchies including those in the \textit{HoloGenOptions}, \textit{HoloGenImageOptions}, \textit{HoloGenProcessOptions},\textit{ HoloGenBatchOptions}, and \textit{SLMControlOptions} libraries. These follow a standard structure with the following classes.
    
    \begin{itemize}
        \item A root element extending \hyperlink{HierarchyRoot}{\texttt{HierarchyRoot}} that provides a name and tool tip as well as containing public properties for all of the \hyperlink{HierarchyPage}{\texttt{HierarchyPage}} objects within. Example: \texttt{OptionsRoot} in the \textit{HoloGenOptions} library.
        
        \item Multiple elements extending \hyperlink{HierarchyPage}{\texttt{HierarchyPage}} that provides a name, tool tip and icon as well as containing public properties for all of the \hyperlink{HierarchyFolder}{\texttt{HierarchyFolder}} objects within. Example: \texttt{ProjectorPage} in the \textit{HoloGenOptions} library.
        
        \item Multiple elements extending \hyperlink{HierarchyFolder}{\texttt{HierarchyFolder}} that provides a name and tool tip as well as containing public properties for all of the \hyperlink{Option}{\texttt{Option}} objects within. Example: \texttt{HologramFolder} in the \textit{HoloGenOptions} library.
        
        \item Multiple elements extending \hyperlink{Option}{\texttt{Option}} or its subclasses that provide the name, tool tip, defaults and limits for the option. Example: \texttt{SLMResolutionX} in the \textit{HoloGenOptions} library. 
        \begin{itemize}
            \item Classes extending \hyperlink{SelectOption}{\texttt{SelectOption}} contain a link to an extension of a \hyperlink{PossibilityCollection}{\texttt{PossibilityCollection}} object. Example: \texttt{SLMTypeOption} in the \textit{HoloGenOptions} library.x
            \item Classes extending \hyperlink{Possibility}{\texttt{Possibility}} or \hyperlink{BooleanOptionWithChildren}{\texttt{BooleanOptionWithChildren}} are also able to contain public properties that will only be editable when selected. Example: \texttt{MultiAmpSLM} in the \textit{HoloGenOptions} library.
        \end{itemize}
        
        \item Elements extending \hyperlink{PossibilityCollection}{\texttt{PossibilityCollection}} with public properties for each of the allowable \hyperlink{Possibility}{\texttt{Possibilities}} of the \hyperlink{SelectOption}{\texttt{SelectOption}}. Example: \texttt{SLMPossibilities} in the \textit{HoloGenOptions} library. 
        
        \item Elements extending \hyperlink{Possibility}{\texttt{Possibility}} with public properties for each of the allowable \hyperlink{Possibility}{\texttt{Possibilities}} of the \hyperlink{SelectOption}{\texttt{SelectOption}}. Example: \texttt{SLMPossibility} in the \textit{HoloGenOptions} library. 
    \end{itemize}
    
    While the reflection based architecture requires an initial investment of time and effort to comes to grips with, it has proved highly time efficient in practice during development on HoloGen. The system presents an alternative to the XML systems commonly used and should be considered for wider use.
    
    \begin{figure}[tbhp]
        \centering
        
            {\includegraphics[width=0.7\linewidth,page=9]{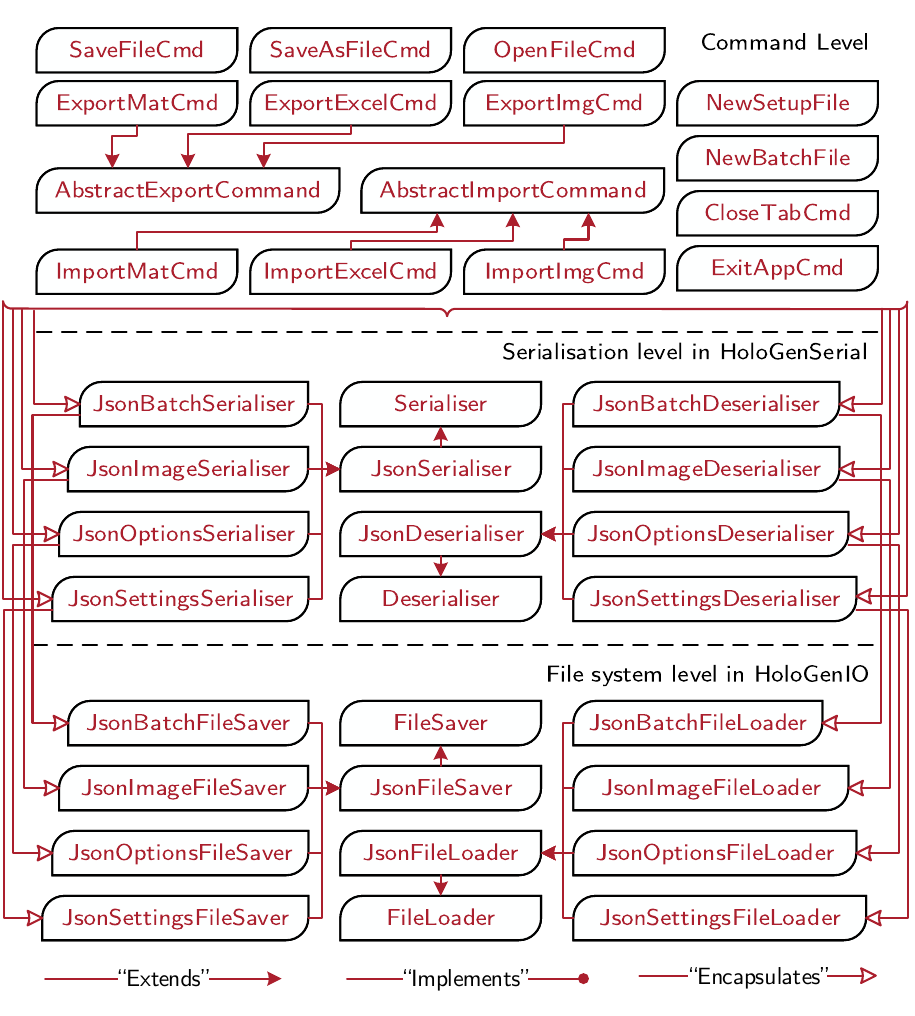}}
        
        \caption[HoloGen Serialisation Architecture]{HoloGen serialisation architecture }
        \label{fig:hierarch4}
    \end{figure}

    \section{Algorithm Interface}\label{codediscuss2}
    
    A four step process is required in order to pass data from the C$\#$ HoloGen application and user interface to the Cuda C/C++ underlying it as shown in Figure~\ref{fig:hierarch3}. 
    
    The top level is the C$\#$ application level where an \texttt{AlgorithmController} unpacks the \hyperlink{Option}{\texttt{Option}} parameter hierarchy and passes it to the Managed C++ level below. All Managed C++ wrapper classes have the suffix "*Man". Managed C++ libraries are able to use both native data types as well as the \textit{.NET} data types used by C$\#$. After the data is copied between the two types, the native C++ layer is called. This level can be exposed in a dynamic or static library but is not able to link to many native Cuda headers such as CUFFT and Thrust. All native C++ wrapper classes have the suffix "*Wrap". These classes can, in turn, pass the data onto Cuda C/C++ compiled classes which are able to communicate with the graphics card. All Cuda C/C++  classes have the suffix "*Cuda".
    
    For example, when passing a target image to a GS algorithm, the following steps occur. \texttt{AlgorithmController} calls \texttt{SetTargetImage()} on a \texttt{GSAlgStandardMan} object it owns. This uses its base \texttt{Convert(GSAlgStandardMan)} functions inherited from \texttt{AlgorithmMan} to transfer the data to native types. This is then passed to the \texttt{GSAlgStandardWrap} and \texttt{GSAlgStandardCuda} objects in turn. Once in the Cuda level, the parent functions in \texttt{AlgorithmCuda} handle transferring the data to the graphics card and maintaining handles on its location. The actual algorithm implementations can be found in the \texttt{RunIterations()} functions of their respective "*Cuda" classes.
    
    While complex, this system performs as fast as direct dll import while allowing for integrated debugging.    
    
    \section{Serialisation}\label{codediscuss3}
    
    Figure~\ref{fig:hierarch4} shows the serialisation architecture for HoloGen. The command level in libraries such as \textit{HoloGenUIMenu} interface to the JSON serialisation and deserialisation classes in the \textit{HoloGenSerial} library. These then write to and read from the file system using the classes found in the \textit{HoloGenIO} library.
    
    \section{User Interface}\label{codediscuss4}
    
    HoloGen follows a standard \textit{View-Model-ViewModel} structure for its user interface. Each UI element or \textit{View} is defined graphically in a *.xaml file, e.g. \texttt{SetupTabView.xaml}, with a C$\#$ companion file, e.g. \texttt{SetupTabView.xaml.cs}. The contained UI elements bind to the data contained and manipulated by the \textit{ViewModel}, e.g. \texttt{SetupTabViewModel.cs}, which in turn holds handles to the internal data, e.g. \texttt{OptionsRoot.cs}. 
    
    \section{Library Descriptions}\label{codediscuss5}
\begin{figure}
        \centering
        
            {\includegraphics[width=0.8\linewidth,page=10]{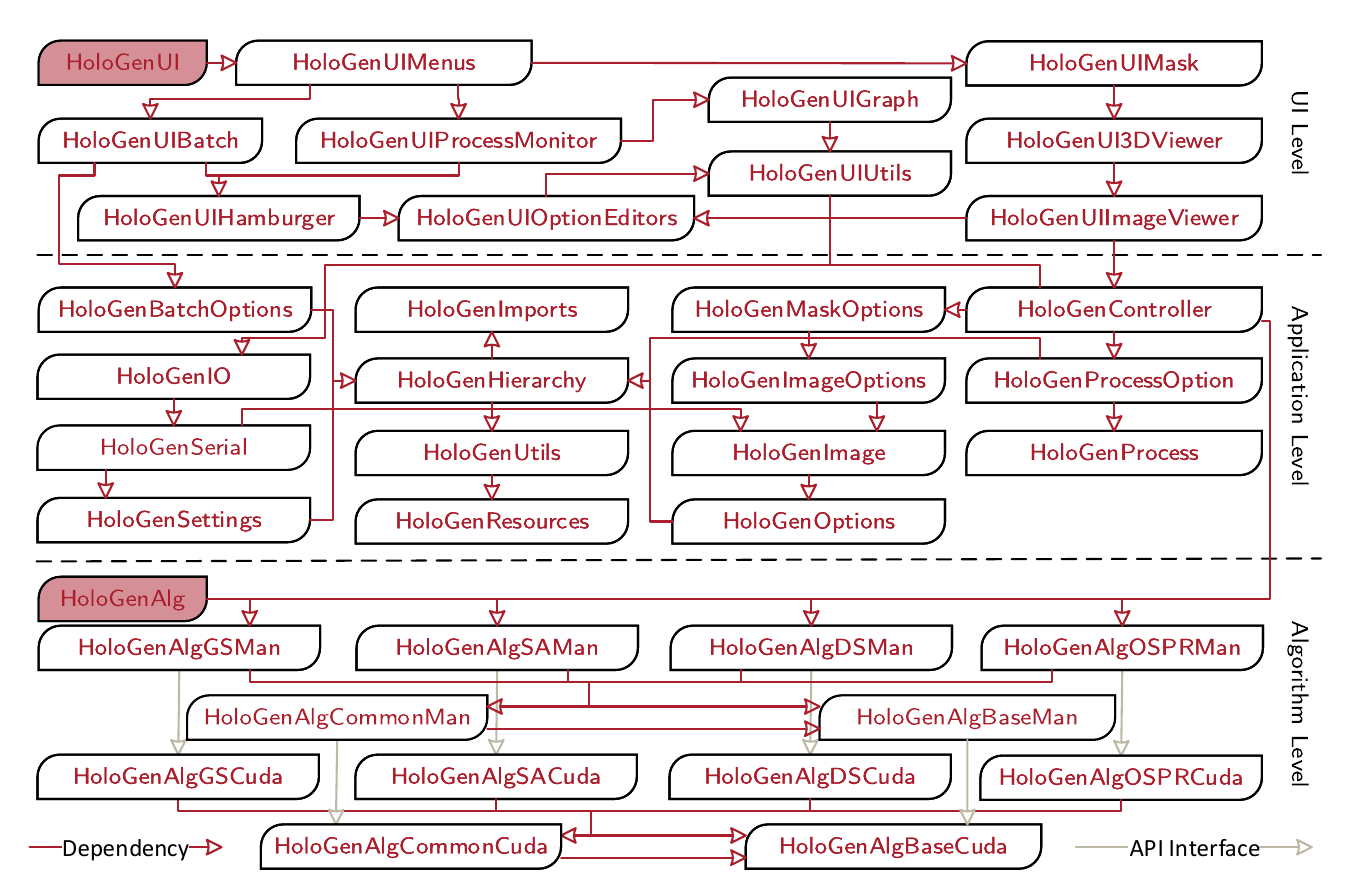}}
        
        \caption[HoloGen Package Layout]{HoloGen package layout. }
        \label{fig:codelayout2d}
    \end{figure}

    Figure~\ref{fig:codelayout2d} shows the layout of libraries within HoloGen with reference to the three application levels shown in Figure~\ref{fig:codelayout1}. This section discusses the libraries used and any key classes not discussed in earlier sections.
    
    \subsection{User Interface Libraries}
    All user interface libraries are prefixed with "HoloGenUI" and are written in C$\#$ using WPF. 
    \begin{itemize}
        \item \hypertarget{HoloGenUI}{\textit{HoloGenUI}} 
        - The main entry point for the user application user interface. Provides top-level implementation \texttt{TabHandlers} for  in \hyperlink{HoloGenUIUtils}{\textit{HoloGenUIUtils}}. Key classes include:
        
        \begin{itemize}
            \item \hypertarget{MainWindowViewModel}{\texttt{MainWindowViewModel}}
            - ViewModel construct underlying the main application window. Keeps an observable collection of \hyperlink{AbstractTab}{\texttt{AbstractTab}} objects and is responsible for matching them with appropriate display classes. Implements \hyperlink{ITabHandler}{\texttt{ITabHandler}} for the \hyperlink{TabHandlerFramework}{\texttt{TabHandlerFramework}}. \hyperlink{MainWindowViewModel}{\texttt{MainWindowViewModel}} is responsible for the application wide user settings including notifications and file IO. Drag and drop behaviours are also overseen by \hyperlink{MainWindowViewModel}{\texttt{MainWindowViewModel}} using the \hyperlink{ControlzEx}{\textit{Dragablz}} library.
            
            \item \hypertarget{AbstractTab}{\texttt{AbstractTab}}
            - Tab data interface that all other tabs can inherit. This allows for templating using  \hyperlink{TabDataTemplateSelector}{\texttt{TabDataTemplateSelector}}.
            
            \item \hypertarget{SetupTabViewModel}{\texttt{SetupTabViewModel}}
            - Concrete implementation of \hyperlink{AbstractTab}{\texttt{AbstractTab}} for setting up a hologram generation process. \hyperlink{SetupTabViewModel}{\texttt{SetupTabViewModel}}'s main function serves to wrap a \hyperlink{HamburgerTabViewModel}{\texttt{HamburgerTabViewModel}} object from \hyperlink{HoloGenUIHamburger}{\textit{HoloGenUIHamburger}}.
            
            \item \hypertarget{BatchTabViewModel}{\texttt{BatchTabViewModel}}
            - Concrete implementation of \hyperlink{AbstractTab}{\texttt{AbstractTab}} for batch processing data. \hyperlink{BatchTabViewModel}{\texttt{BatchTabViewModel}}'s main function serves to wrap a \hyperlink{BatchViewModel}{\texttt{BatchViewModel}} object from \hyperlink{HoloGenUIBatch}{\textit{HoloGenUIBatch}}.
            
            \item \hypertarget{BrowserTabViewModel}{\texttt{BrowserTabViewModel}}
            - Concrete implementation of \hyperlink{AbstractTab}{\texttt{AbstractTab}} providing a built in browser. \hyperlink{BrowserTabViewModel}{\texttt{BrowserTabViewModel}}'s main function serves to wrap the imported \hyperlink{CEFSharp}{\textit{CEFSharp}} functionality.
            
            \item \hypertarget{ProcessTabViewModel}{\texttt{ProcessTabViewModel}}
            - Concrete implementation of \hyperlink{AbstractTab}{\texttt{AbstractTab}} for visualising running process parameters and reporting messages from the underlying algorithms. \hyperlink{ProcessTabViewModel}{\texttt{ProcessTabViewModel}}'s main function serves to wrap a \hyperlink{ProcessViewModel}{\texttt{ProcessViewModel}} object from \hyperlink{HoloGenUIProcessMonitor}{\textit{HoloGenUIProcessMonitor}}.
            
            \item \hypertarget{HologramTabViewModel}{\texttt{HologramTabViewModel}}
            - Concrete implementation of \hyperlink{AbstractTab}{\texttt{AbstractTab}} for visualising complex valued images in 2d and 3D with optional masking data.. \hyperlink{HologramTabViewModel}{\texttt{HologramTabViewModel}}'s main function serves to wrap a \hyperlink{MaskViewModel}{\texttt{MaskViewModel}} object from \hyperlink{HoloGenUIMask}{\textit{HoloGenUIMask}}.
            
        \end{itemize}
        
        The display code is written in C$\#$ and WPF and lies on top of the ViewModels. There is a one to one correspondence between ViewModel classes and XAML parameter sheets on top.
        
        \item \hypertarget{HoloGenUIMenus}{\textit{HoloGenUIMenus}} 
        - Defines \hyperlink{Options}{\texttt{Options}} and \hyperlink{Commands}{\texttt{Commands}} for the main application flyout menu. Built on-top of the \hyperlink{HoloGenHierarchy}{\textit{HoloGenHierarchy}} library discussed in Section~\ref{codediscuss1}.
        
        \item \hypertarget{HoloGenUIHamburger}{\textit{HoloGenUIHamburger}} 
        - Defines a series of View/ViewModel pairs that serve to unpack and display \hyperlink{HoloGenHierarchy}{\textit{HoloGenHierarchy}} parameter hierarchies in a Microsoft Metro "Hamburger Menu" style. Individual editors are defined in \hyperlink{HoloGenUIOptionEditors}{\textit{HoloGenUIOptionEditors}}.
        
        \item \hypertarget{HoloGenUIOptionEditors}{\textit{HoloGenUIOptionEditors}} 
        - Defines a series of View/ViewModel pairs that serve to display the different \hyperlink{Option}{\textit{Option}} types discussed in Section~\ref{codediscuss1}. Primarily used as the \textit{leaf nodes} for \hyperlink{HoloGenUIHamburger}{\textit{HoloGenUIHamburger}}.
        
        \item \hypertarget{HoloGenUIProcessMonitor}{\textit{HoloGenUIProcessMonitor}} 
        - Defines display classes for the algorithm monitoring tab. The charting is built on top of \hyperlink{LiveCharts}{\textit{LiveCharts}} using \hyperlink{HoloGenUIGraph}{\textit{HoloGenUIGraph}}.
        
        \item \hypertarget{HoloGenUIGraph}{\textit{HoloGenUIGraph}} 
        - Defines display classes for realtime charting. \hyperlink{HoloGenUIGraph}{\textit{HoloGenUIGraph}} is built on top of \hyperlink{LiveCharts}{\textit{LiveCharts}}.
        
        \item \hypertarget{HoloGenUIResources}{\textit{HoloGenUIResources}} 
        - Defines translation strings for the application localisation. 
        
        \item \hypertarget{HoloGenUIBatch}{\textit{HoloGenUIBatch}} 
        - Defines display classes for the batch processing tab. Defines a number of \hyperlink{AbstractColumnFactory}{\textit{AbstractColumnFactory}} implementations for displaying different \hyperlink{Option}{\texttt{Option}} types. Key classes include:
        
        \begin{itemize}
            \item \hypertarget{AbstractColumnFactory}{\texttt{AbstractColumnFactory}} 
            - Interface for a series of factory objects - \texttt{BooleanColumnFactory}, \texttt{DefaultColumnFactory}, \texttt{DoubleColumnFactory}, \texttt{IntegerColumnFactory}, \texttt{PathColumnFactory}, \texttt{SelectColumnFactory}, and \texttt{TextColumnFactory} - used for generating WPF \texttt{DataGridColumns}.
            
            \item \hypertarget{TemplateGenerator}{\texttt{TemplateGenerator}} 
            - Helper class for \hyperlink{AbstractColumnFactory}{\texttt{AbstractColumnFactory}} implementations that uses a given delegate to create new instances for similar \hyperlink{Option}{\texttt{Option}} types.
        \end{itemize}
        
        \item \hypertarget{HoloGenUIImageViewer}{\textit{HoloGenUIImageViewer}} 
        - Defines classes for viewing complex valued images in 2D.
        
        \item \hypertarget{HoloGenUIMask}{\textit{HoloGenUIMask}} 
        - Extends the functionality from \hyperlink{HoloGenUIImageViewer}{\textit{HoloGenUIImageViewer}} showing masked regions on a 2D complex valued image.
        
        \item \hypertarget{HoloGenUI3DViewer}{\textit{HoloGenUI3DViewer}} 
        - Extends the functionality from \hyperlink{HoloGenUIMask}{\textit{HoloGenUIMask}} showing complex valued images in 3D.
        
        \item \hypertarget{HoloGenUIUtils}{\textit{HoloGenUIUtils}} 
        - Defines low level utility classes for the HoloGen UI as well as abstract interfaces for services offered by higher level libraries.
        Key classes include:
        \begin{itemize}
            \item \hypertarget{ITabHandler}{\texttt{ITabHandler}} and \hypertarget{TabHandlerFramework}{\texttt{TabHandlerFramework}}
            - Define a service framework. The \hyperlink{MainWindowViewModel}{\texttt{MainWindowViewModel}} extends \hyperlink{ITabHandler}{\texttt{ITabHandler}} and can register itself with \hyperlink{TabHandlerFramework}{\texttt{TabHandlerFramework}}. Lower level class libraries can then call the \hyperlink{TabHandlerFramework}{\texttt{TabHandlerFramework}} while remaining ignorant of the implementation.
            
            \item \hypertarget{CreatorsThesis}{\texttt{CreatorsThesis}} 
            - Display theme for the application built on top of \hyperlink{MahApps}{\textit{MahApps}}.
            
            \item \hypertarget{ICanExport}{\texttt{ICanExport}}, \hypertarget{ICanExportBitmap}{\texttt{ICanExportBitmap}}, \hypertarget{ICanExportExcel}{\texttt{ICanExportExcel}}, \hypertarget{ICanExportMat}{\texttt{ICanExportMat}} and \hypertarget{ICanSave}{\texttt{ICanSave}} 
            - Interfaces that \hyperlink{ITabHandler}{\texttt{ITabHandler}} implementations can extend to declare their IO requirements.
        \end{itemize}
    \end{itemize}
    \subsection{Application Libraries}
    All application libraries are prefixed with "HoloGen".
    \begin{itemize}
        
        \item \hypertarget{HoloGenController}{\textit{HoloGenController}} 
        - Defines classes that unwrap the \hyperlink{Options}{\texttt{Options}} hierarchy defined in \hyperlink{HoloGenOptions}{\textit{HoloGenOptions}} for hologram generation and communicate it with the algorithms wrapped by \hyperlink{HoloGenAlgBaseMan}{\textit{HoloGenAlgBaseMan}}, \hyperlink{HoloGenAlgGSMan}{\textit{HoloGenAlgGSMan}}, \hyperlink{HoloGenAlgSAMan}{\textit{HoloGenAlgSAMan}}, \hyperlink{HoloGenDSMan}{\textit{HoloGenDSMan}} and \hyperlink{HoloGenOSPRMan}{\textit{HoloGenOSPRMan}}. Key class is \hyperlink{AlgorithmController}{\texttt{AlgorithmController}}.
        
        \item \hypertarget{HoloGenProcess}{\textit{HoloGenProcess}} 
        - Defines data structures for defining a hologram generation process.
        
        \item \hypertarget{HoloGenImage}{\textit{HoloGenImage}} 
        - Defines classes related to complex valued images.
        Key classes include:
        \begin{itemize}
            \item \hypertarget{ComplexImage}{\texttt{ComplexImage}}
            - Object that holds an image in \texttt{Complex} format as well as generation metadata and pre-cached values. Defines specialised JSON interface commands in order to preserve disk space.
            
            \item \hypertarget{ImageCache}{\texttt{ImageCache}}
            - Defines a cache for different \texttt{Bitmap} views on a \hyperlink{ComplexImage}{\texttt{ComplexImage}} object. Once a particular visualisation bitmap is generated for a particular image, the result is cached to reduce future load times. Uses a \hyperlink{GenericCache}{\texttt{GenericCache}} from \hyperlink{HoloGenUtils}{\textit{HoloGenUtils}} as the underlying implementation with  the \texttt{TransformType}, \texttt{ImageViewType}, \texttt{ColorScheme} and \texttt{ImageScaleType} enums as the four access keys.
        \end{itemize}
        
        \item \hypertarget{HoloGenHierarchy}{\textit{HoloGenHierarchy}} 
        - Defines the \hyperlink{Options}{\texttt{Options}} and \hyperlink{Command}{\texttt{Command}} hierarchy discussed in Section~\ref{codediscuss1}.
        
        \item \hypertarget{HoloGenImports}{\textit{HoloGenImports}} 
        - Classes or constructs that have been imported from other applications in a source code format.
        
        \item \hypertarget{HoloGenResources}{\textit{HoloGenResources}} 
        - Defines translation strings for the application localisation. 
        
        \item \hypertarget{HoloGenOptions}{\textit{HoloGenOptions}} 
        - Defines an \hyperlink{Options}{\texttt{Options}} hierarchy for the hologram generation algorithms. Built on-top of the \hyperlink{HoloGenHierarchy}{\textit{HoloGenHierarchy}} library discussed in Section~\ref{codediscuss1}.
        
        \item \hypertarget{HoloGenProcessOptions}{\textit{HoloGenProcessOptions}} 
        - Defines an \hyperlink{Options}{\texttt{Options}} hierarchy for manipulating the running process display in \hyperlink{HoloGenUIProcessMonitor}{\textit{HoloGenUIProcessMonitor}}. Built on-top of the \hyperlink{HoloGenHierarchy}{\textit{HoloGenHierarchy}} library discussed in Section~\ref{codediscuss1}.
        
        \item \hypertarget{HoloGenBatchOptions}{\textit{HoloGenBatchOptions}} 
        - Defines an \hyperlink{Options}{\texttt{Options}} hierarchy for the batch processing hologram generation algorithms in \hyperlink{HoloGenUIBatch}{\textit{HoloGenUIBatch}}. Built on-top of the \hyperlink{HoloGenHierarchy}{\textit{HoloGenHierarchy}} library discussed in Section~\ref{codediscuss1}.
        
        \item \hypertarget{HoloGenSettings}{\textit{HoloGenSettings}} 
        - Defines an \hyperlink{Options}{\texttt{Options}} hierarchy for the application settings in \hyperlink{HoloGenUI}{\textit{HoloGenUI}} and \hyperlink{HoloGenUIMenus}{\textit{HoloGenUIMenus}}. Built on-top of the \hyperlink{HoloGenHierarchy}{\textit{HoloGenHierarchy}} library discussed in Section~\ref{codediscuss1}.
        
        \item \hypertarget{HoloGenImageOptions}{\textit{HoloGenImageOptions}} 
        - Defines an \hyperlink{Options}{\texttt{Options}} hierarchy for visualising complex valued images in \hyperlink{HoloGenUIImageViewer}{\textit{HoloGenUIImageViewer}}. Built on-top of the \hyperlink{HoloGenHierarchy}{\textit{HoloGenHierarchy}} library discussed in Section~\ref{codediscuss1}.
        
        \item \hypertarget{HoloGenMaskOptions}{\textit{HoloGenMaskOptions}} 
        - Defines an \hyperlink{Options}{\texttt{Options}} hierarchy for visualising complex valued images with masking data in \hyperlink{HoloGenUIMask}{\textit{HoloGenUIMask}}. Built on-top of the \hyperlink{HoloGenHierarchy}{\textit{HoloGenHierarchy}} library discussed in Section~\ref{codediscuss1}.
        
        \item \hypertarget{HoloGenSerial}{\textit{HoloGenSerial}} 
        - Handles the serialisation of \hyperlink{HierarchySaveable}{\texttt{HierarchySaveable}} data objects. Discussed further in Section~\ref{codediscuss4}.
        
        \item \hypertarget{HoloGenIO}{\textit{HoloGenIO}} 
        - Handles the file input an doutput of serialised of \hyperlink{HierarchySaveable}{\texttt{HierarchySaveable}} data objects. Discussed further in Section~\ref{codediscuss4}.
        
        \item \hypertarget{HoloGenUtils}{\textit{HoloGenUtils}} 
        - Utility classes for HoloGen at the application level.
        Key classes include:
        \begin{itemize}
            \item \hypertarget{ComplexImage}{\texttt{ComplexImage}}
            - Object that holds an image in \texttt{Complex} format as well as generation metadata and pre-cached values. Defines specialised JSON interface commands in order to preserve disk space.
            
            \item \hypertarget{GenericCache}{\texttt{GenericCache}}
            - Defines a cache for computationally expensive results using up to four different keys. Lambda functions are used for the computation to increase reusability. Provides the base implementation for \hyperlink{ImageCache}{\texttt{ImageCache}}. 
        \end{itemize}
        
        \item \hypertarget{HoloGen}{\textit{HoloGen}} 
        - Command line interface for the HoloGen application.
    \end{itemize}
    
    \subsection{Algorithm Libraries}
    All algorithm libraries are prefixed with "HoloGenAlg" and recieve the "Man" or "Cuda" suffix depending on whether they are written in Managed C++ or Cuda C/C++.
    \begin{itemize}
        \item \hypertarget{HoloGenAlg}{\textit{HoloGenAlg}} 
        - Command line interface for the HoloGen algorithms. Used for batch processing test data in a client-server configuration.
        
        \item \hypertarget{HoloGenAlgBaseMan}{\textit{HoloGenAlgBaseMan}} 
        - Managed C++ wrapper for \hyperlink{HoloGenAlgBaseCuda}{\textit{HoloGenAlgBaseCuda}}.
        
        \item \hypertarget{HoloGenAlgCommonMan}{\textit{HoloGenAlgCommonMan}} 
        - Managed C++ wrapper for \hyperlink{HoloGenAlgCommonCuda}{\textit{HoloGenAlgCommonCuda}}.
        
        \item \hypertarget{HoloGenAlgGSMan}{\textit{HoloGenAlgGSMan}} 
        - Managed C++ wrapper for \hyperlink{HoloGenAlgGSCuda}{\textit{HoloGenAlgGSCuda}}.
        
        \item \hypertarget{HoloGenAlgSAMan}{\textit{HoloGenAlgSAMan}} 
        - Managed C++ wrapper for \hyperlink{HoloGenAlgSACuda}{\textit{HoloGenAlgSACuda}}.
        
        \item \hypertarget{HoloGenAlgDSMan}{\textit{HoloGenAlgDSMan}} 
        - Managed C++ wrapper for \hyperlink{HoloGenAlgDSCuda}{\textit{HoloGenAlgDSCuda}}.
        
        \item \hypertarget{HoloGenAlgOSPRMan}{\textit{HoloGenAlgOSPRMan}} 
        - Managed C++ wrapper for \hyperlink{HoloGenAlgOSPRCuda}{\textit{HoloGenAlgOSPRCuda}}.
        
        \item \hypertarget{HoloGenAlgBaseCuda}{\textit{HoloGenAlgBaseCuda}} 
        - C++ and Cuda C base level algorithm definitions. Discussed further in Section~\ref{codediscuss2}. 
        Key classes include:
        \begin{itemize}
            \item \hypertarget{AlgorithmCuda}{\texttt{AlgorithmCuda}}
            - Base level algorithm definition. Stores target, illumination and starting images as well as common algorithm parameters.
            
            \item \hypertarget{FFTHandlerCuda}{\texttt{FFTHandlerCuda}}
            - Wraps the \textit{CUFFT} FFT library with \textit{Thrust} friendly functions.
            
            \item \hypertarget{FFTUpdaterCuda}{\texttt{FFTUpdaterCuda}}
            - Updates the replay field after the change of a single refraction field pixel.
            
            \item \hypertarget{Normaliser}{\texttt{Normaliser}}
            - Provides a fast vector normalisation feature.
            
            \item \hypertarget{Randomiser}{\texttt{Randomiser}}
            - Allows for randomisation of different combinations of phase and amplitude.
            
            \item \hypertarget{QuantiserCuda}{\texttt{QuantiserCuda}}
            - Base level quantisation algorithm.
            
            \item \hypertarget{StridedChunkRange}{\texttt{StridedChunkRange}}
            - Customised \textit{Thrust} iterator that can iterate through square regions of a matrix.
        \end{itemize}
        
        \item \hypertarget{HoloGenAlgCommonCuda}{\textit{HoloGenAlgCommonCuda}} 
        - C++ and Cuda C common utility functions.
        
        \item \hypertarget{HoloGenAlgGSCuda}{\textit{HoloGenAlgGSCuda}} 
        - C++ and Cuda C implementation of the Gerchberg-Saxton algorithm and other iterative Fourier transform approaches. 
        
        \item \hypertarget{HoloGenAlgSACuda}{\textit{HoloGenAlgSACuda}} 
        - C++ and Cuda C implementation of the simulated annealing algorithm. 
        
        \item \hypertarget{HoloGenAlgDSCuda}{\textit{HoloGenAlgDSCuda}} 
        - C++ and Cuda C implementation of the direct binary search algorithm. 
        
        \item \hypertarget{HoloGenAlgOSPRCuda}{\textit{HoloGenAlgOSPRCuda}} 
        - C++ and Cuda C implementation of the One-Step Phase-Retrieval algorithm. 
    \end{itemize}

    \subsection{Imported Libraries}
    HoloGen also uses a number of imported libraries. A brief description of their function is presented here.
    \begin{itemize}
        \item \hypertarget{MahApps}{\textit{MahApps}}
        - Custom controls for WPF apps as well as a material design skin based on Microsoft's Metro UI.
        
        \item \hypertarget{LiveCharts}{\textit{LiveCharts}}
        - Customisable and bindable real-time charting library.
        
        \item \hypertarget{Dragablz}{\textit{Dragablz}}
        - Draggable tabs for WPF.
        
        \item \hypertarget{ControlzEx}{\textit{ControlzEx}}
        - Custom controls for WPF apps.
        
        \item \hypertarget{HelixToolkit}{\textit{HelixToolkit}}
        - 3D viewer for WPF apps.
        
        \item \hypertarget{Xamarin.Forms}{\textit{Xamarin.Forms}}
        - Mobile/tablet compatibility.
        
        \item \hypertarget{MaterialDesign}{\textit{MaterialDesign}}
        - Material Design compatible skin for WPF apps.
        
        \item \hypertarget{MaterialSkin}{\textit{MaterialSkin}}
        - Alternative Material Design compatible skin for WPF apps.
        
        \item \hypertarget{GongSolutions.WPF.DragDrop}{\textit{GongSolutions.WPF.DragDrop}}
        - WPF drag and drop capability.
        
        \item \hypertarget{FastMember.Signed}{\textit{FastMember.Signed}}
        - Fast reflection for \textit{.NET}.
        
        \item \hypertarget{DocumentFormat.OpenXML}{\textit{DocumentFormat.OpenXML}}
        - Microsoft Office file format interoperability.
        
        \item \hypertarget{ExcelNumberFormat}{\textit{ExcelNumberFormat}}
        - Advanced Excel number formatting.
        
        \item \hypertarget{SharpDX}{\textit{SharpDX}}
        - WPF DirectX compatibility. 
        
        \item \hypertarget{ClosedXML}{\textit{ClosedXML}}
        - Microsoft Excel integration.
        
        \item \hypertarget{CefSharp}{\textit{CefSharp}}
        - WPF compatibly wrapper for the Chromium browser.
        
        \item \hypertarget{Cuda}{\textit{Cuda}}
        - Programmable interface to NVidia graphics cards.
        
        \item \hypertarget{PdfiumViewer}{\textit{PdfiumViewer}}
        - PDF file viewer.
        
        \item \hypertarget{Xceed}{\textit{Xceed}}
        - DataGrid controls for WPF apps.
        
        \item \hypertarget{NHotKey}{\textit{NHotKey}}
        - Global hot keys for WPF apps.
        
        \item \hypertarget{Newtonsoft.Json}{\textit{Newtonsoft.Json}}
        - JSON serialisation for .NET languages.
        
        \item \hypertarget{MathNet}{\textit{MathNet}}
        - Mathematics package for .NET languages.
        
        \item \hypertarget{AForge}{\textit{AForge}}
        - Mathematics package for .NET languages.
        
        \item \hypertarget{Accord}{\textit{Accord}}
        - Mathematics package for .NET languages.
        
        \item \hypertarget{NUnit}{\textit{NUnit}}
        - unit test framework for .NET languages.
    \end{itemize}

    \section{Further Research and Conclusion}\label{codediscuss6}
    
    This article has presented a brief summary of the structure and design of HoloGen, an open-source hologram generation package. While HoloGen is feature rich, it is by no means a complete package. In particular, the number of algorithms available is limited and additional algorithms should be implemented. 
    
    Development of HoloGen is ongoing and the authors welcome any feedback, advice and assistance offered during this process.
    
    \normalsize
    \bibliography{references}
    
    
\end{document}